\documentclass[12pt,a4paper]{article}
\usepackage{latexsym}
\usepackage{graphicx}

\def\de{\partial}

\def\K1{{\cal K}_{\bf 1}}
\def\Q20{{\cal Q}_{\bf 20'}}

\newcommand{\tr}{{\rm tr}}
\newcommand{\ie}{{\em i.e.~}}
\newcommand{\eg}{{\em e.g.~}}
\newcommand{\be}{\begin{equation}}
\newcommand{\ee}{\end{equation}}
\newcommand{\ba}{\begin{eqnarray}}
\newcommand{\ea}{\end{eqnarray}}
\newcommand{\polylog} {{\rm Li}}
\begin{document}

\begin{titlepage}
\thispagestyle{empty}

\begin{flushright}
ROM2F/2008/18 \\
%DAMTP-2000-34
\end{flushright}

\vspace{1.5cm}

\begin{center}

{\LARGE {\bf The Massless Supersymmetric Ladder with $L$ Rungs\rule{0pt}{25pt} }} \\
\vspace{1cm} \ { G.~C.~Rossi$^{\dagger}$
and Ya.~S.~Stanev$^{\ddag}$ } \\
\vspace{0.6cm} {
{\it Dipartimento di Fisica, \ Universit\`a di Roma \ ``Tor Vergata''}} \\  {{\it I.N.F.N.\ -\ Sezione
di Roma \ ``Tor Vergata''}} \\ {{\it Via della Ricerca  Scientifica, 1}}
\\ {{\it 00133 \ Roma, \ ITALY}}
\end{center}

\vspace{1cm}

\begin{abstract}

We show that in the massless $N=1$ supersymmetric Wess--Zumino  theory it is possible to devise a computational
strategy by which the $x$-space calculation of the ladder 4-point correlators can be carried out without
introducing any regularization. As an application we derive a representation valid  at all loop
orders  in terms of conformal invariant integrals. We obtain an explicit expression of the 3-loop
ladder diagram for collinear external points.

\end{abstract}

\vspace{4cm}
\noindent
\rule{6.5cm}{0.4pt}

\begin{flushleft}
$^{\dagger}$ Giancarlo.Rossi@roma2.infn.it

$^{\ddag}$ Yassen.Stanev@roma2.infn.it
\end{flushleft}

\end{titlepage}

\vfill
\newpage

\section{Introduction}
\label{sec:INTRO}

Supersymmetry has many remarkable properties. One of them is that in general
supersymmetric theories are less divergent than non-supersymmetric ones~\cite{WZ,norenorm1,norenorm2,GRS}
(for a review see~\cite{1001,west_book}).
This property, however, is not always easy
to use, since typically one has to express the finite super Feynman diagrams in terms
of ordinary (coordinate or momentum space) integrals which require regularization.

In this paper we show that it is possible to considerably simplify the expression of the
supersymmetric massless ladder 4-point diagrams  at
arbitrary loop order, and rewrite them as 4-point conformal integrals.
Interestingly we are able to carry out the whole calculation without introducing
any regulator, despite the fact that individual component diagrams are divergent.
Before going into details, let us briefly sketch our approach.

Consider a certain $N=1$ superdiagram, which by super power-counting is finite, for example
the irreducible massless 4-point ladder diagram with four external scalar legs~\footnote{
In this paper we shall concentrate mostly on this case,
but similar considerations hold also for other finite superdiagrams.} depicted in Figure~\ref{fig:fig1}.
Although the expression for the non supersymmetric massless $\varphi^3$ ladder with an arbitrary number of rungs
was found long ago in~\cite{davyd2}, its generalization to the supersymmetric case is still not known.

To compute the ladder diagram in Figure~\ref{fig:fig1} at a given order in perturbation theory, in coordinate space,
one has to evaluate a multiple superspace integral of the form
\ba
G & \equiv &  G(x_1,x_2,x_3,x_4) = \nonumber \\
 & = & \int \prod_k  d^4y_k \; d^2 \theta_k \; d^2 \bar \theta_k
\; {\cal W} (x_1,x_2,x_3,x_4,\{y_k\},\{\theta_k\},\{\bar \theta_k\}) \ ,
\label{sd1}
\ea
where ${\cal W} (x_1,x_2,x_3,x_4,\{y_k\},\{\theta_k\},\{\bar \theta_k\})$
denotes the product of super propagators, deriving from
 Wick contractions. Performing  $\theta$ and $\bar \theta$
integrations in general gives as a result a linear combination of several component Feynman diagrams,
so we can write eq.~(\ref{sd1}) in the form
\be
G =  \int \prod_k d^4y_k
\sum_i { W_i} (x_1,x_2,x_3,x_4,\{y_k\}) \ ,
\label{sd2}
\ee
where ${ W_i} (x_1,x_2,x_3,x_4,\{y_k\})$ denote all allowed  products of component field propagators
 (diagrams).
The standard procedure for computing $G$
is to exchange in eq.~(\ref{sd2}) the order in which the sum over $i$
and the integration over $\{y_k\}$
are performed,  obtaining
\be
\widetilde {G} =  \sum_i  \int \prod_k  d^4y_k
{ W_i} (x_1,x_2,x_3,x_4,\{y_k\}) \ .
\label{sd3}
\ee
Note, however, that although by assumption the function $G$ is finite,
some of the terms in $\widetilde {G}$ may diverge, so one has to
introduce a regularization prescription to compute them.
The regulator can be removed only after summing up all the terms.

This suggests that the necessity of regularization is not an intrinsic
feature of finite correlation functions, but an artefact of our way of representing them
in terms of ordinary component Feynman diagrams.
Comparing eq.~(\ref{sd2}) and eq.~(\ref{sd3}) it is evident that the problem arises from
exchanging  the order of  the integrations over $\{y_k\}$ and the summation over $i$.
As we shall show, at least for the
class of massless ladder 4-point superdiagrams, one can follow a different approach, which
does not require any regulator at  intermediate steps.
To be explicit, we allow ourselves to exchange the order in which the integrations and the sum
are performed in eq.~(\ref{sd2}) only as far as the result
for each individual contribution to the sum remains finite. We then compute the integrals under the
sum, perform the summation and then evaluate the remaining integrals. In formulae
this corresponds to the following representation
\be
{G} = \int \prod_{k_1} d^4y_{k_1} \sum_i  \int \prod_{ k_2} d^4y_{k_2}
{ W_i} (x_1,x_2,x_3,x_4,\{y_k\}) \ ,
\label{sd4}
\ee
where each of the integrals
$$
\int \prod_{ k_2} d^4y_{k_2} { W_i} (x_1,x_2,x_3,x_4,\{y_k\})
$$
is finite.
The splitting of the integration points $\{y_k\}$ into two sets $\{y_{k_1}\}$ and $\{y_{k_2}\}$
depends on the diagram under consideration and, as we will show, in general is not unique.
Different choices may  give rise to
apparently very different representations for the same correlation function $G$.

The paper is organized as follows. In Section~\ref{sec:LSD} we set up the problem and
provide a simplified expression of
the 3-loop supersymmetric ladder diagram.
In Section~\ref{sec:AOR}  we extend this result to all loop orders
and express the massless supersymmetric ladder diagram with $L$ rungs in terms of conformal
integrals.
In Section~\ref{sec:ARG} we derive two different diagrammatic representations for these conformal integrals.
In Section~\ref{sec:3LP}, as a first step towards the evaluation of these integrals,
we obtain an explicit expression
for the 3-loop amplitude in the special case when all the points lie on a straight line
and compute the singularities of the function. Finally,
 in Section~\ref{sec:CONO} we give some conclusions.

\section{The ladder super diagram}
\label{sec:LSD}

We consider the
massless Wess--Zumino model~\cite{WZ} with (Euclidean space) action
\ba
S &=& \int d^4 x  \left\{  \int d^2 \theta d^2 \bar \theta \;
\Phi^{\dagger}(x,\theta ,\bar \theta) \, \Phi(x,\theta,\bar \theta)
\right.
\nonumber \\
& - & \left. {g \over 3!} \int d^2 \theta \, \left(\Phi(x,\theta,0)\right)^3 -
{g \over 3!} \int d^2 \bar \theta  \, \left(\Phi^{\dagger}(x,0 ,\bar \theta)\right)^3
\right \} \ .
\label{action}
\ea
The chiral $N=1$ superfield $\Phi(x,\theta,\bar \theta)$
contains a scalar $\varphi$, a Weyl fermion $\psi$ and an auxiliary filed $F$.
The $\langle \Phi^{\dagger} \;\Phi \rangle$ propagator can be written in the compact form
\be
\langle \Phi^{\dagger}(x_i,\theta_i,\bar \theta_i) \;
\Phi(x_j,\theta_j ,\bar \theta_j) \rangle = {1 \over 4 \pi^2}\;
{\rm e}^{(\xi_{ii}+\xi_{jj}-2\xi_{ji} ) . \de_j} {1 \over x_{ij}^2} \ ,
\label{susypropagator}
\ee
where
$x_{ij} = x_i-x_j$, $\de_j = \de / \de{x^j}$ and
$ \xi^{\mu}_{ij}= \theta_i^\alpha \sigma^{\mu}_{\alpha \dot \beta} \bar \theta_j^{\dot \beta}$.
In this model only the 2-point function is divergent, while all  higher point functions are
finite. We remind  that in the massless case the 3-point function
may receive finite corrections~\cite{west,JJwest}.

\begin{figure} [ht]
%[!htb]
% [!htbp]
 %\begin{minipage}[t]{\linewidth}

  \vskip -90pt
    \centering
    \includegraphics[width=1\linewidth]{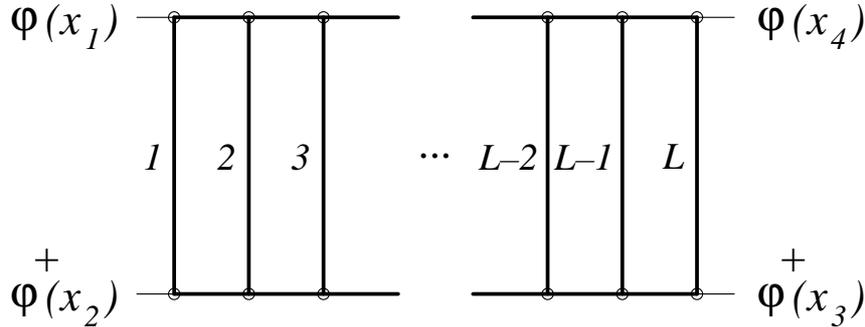}
  \vskip -60pt
  %   \vskip -8cm
\caption{The ladder diagram with $L$ rungs.}
  \label{fig:fig1}
 %\end{minipage} \\ [2.5cm]
\end{figure}

\begin{figure} [ht]
%[!htb]
% [!htbp]
 %\begin{minipage}[t]{\linewidth}

  \vskip -40pt
    \centering
    \includegraphics[width=1\linewidth]{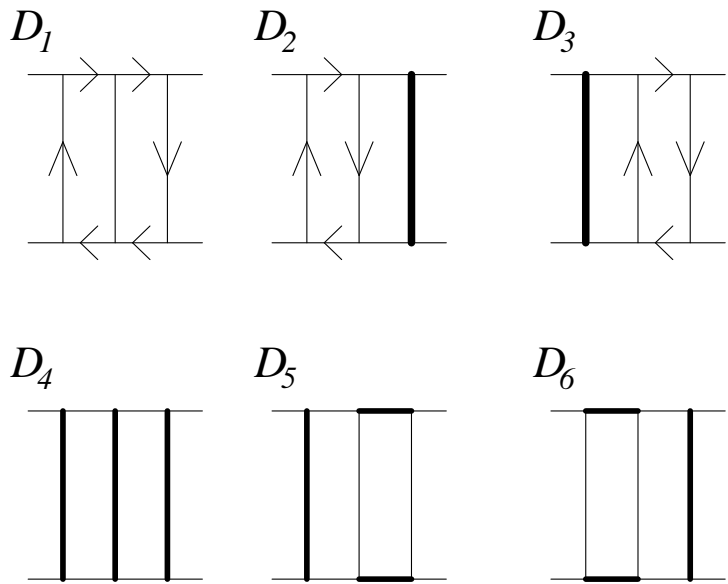}
  \vskip -40pt
  %   \vskip -8cm
\caption{The six component diagrams for $L=3$.}
  \label{fig:fig2}
 %\end{minipage} \\ [2.5cm]
\end{figure}

In this paper we shall concentrate
on the irreducible massless ladder superdiagram with four external scalar legs $\varphi$ and $L$ rungs,
which is depicted in Figure~\ref{fig:fig1}, namely
\be
G_L(x_1,x_2,x_3,x_4) = \langle \varphi (x_1) \varphi^{\dagger} (x_2) \varphi^{\dagger} (x_3) \varphi (x_4) \rangle
|_{g^{2L}, \; {\rm ladder}} \, .
\label{G_L}
\ee
$G_L$ is proportional to $g^{2L}$, so in the standard loop-order counting it
is a  $L$ loop diagram.
%%%%%%%%%%%%%%%%%%%%%%%
Let us stress that with our choice for the external fields, the ladder in Figure~\ref{fig:fig1} is planar
for odd values of $L$, while for even values of $L$ it is twisted.
In Figure~\ref{fig:fig1} all the internal lines represent superpropagators, hence effectively this single
picture corresponds to a set of component field diagrams. In Figure~\ref{fig:fig2} we present as an example
the 6 component diagrams corresponding to the $L=3$ superdiagram.
%%%%%%%%%%%%%%%%%%%%%%%

Here and later in the paper for simplicity we
shall suppress systematically all the factors $4 \pi^2$ in the propagators.
Indeed, since $G_L$ involves $3 L +2$ propagators, the overall factor
$1/(4 \pi^2)^{3L+2}$ can be reinstated at the end of the calculation.
We  use the following graphical notation for the
component field propagators.
We denote the $\varphi$ propagator, $ 1/x^2$, by a thin continuous line,
the fermion $\psi$ propagator, $\hat x / x^4 = x_{\mu} \sigma^{\mu} /x^4$, by a thin continuous line with an arrow, and we
 represent by a fat line the  $\delta$-function, corresponding to the propagator of the auxiliary field $F$.
From the superpropagator of eq.~(\ref{susypropagator}) one can compute the relative sign factors weighting the
different component diagrams in a superdiagram. It turns out that there is a $(-1)$ for each
closed fermionic loop. Note also  that with our graphical conventions there is
a $(-1)$ sign for each $F$ propagator, since  in Euclidean space the $F$ propagator is
\be \Box {1 \over x^2} = - 4 \pi^2  \; \delta(x) \ .
\label{box}
\ee
With this conventions one finds that the expansion of the superdiagram in Figure~\ref{fig:fig1},  for $L=3$, in
terms of the diagrams in Figure~\ref{fig:fig2} is
\be
G_3(x_1,x_2,x_3,x_4) = -D_1+D_2+D_3-D_4-D_5-D_6 \ .
\label{G_3}
\ee
Note that while the superdiagram in Figure~\ref{fig:fig1} is finite by $N=1$ super power counting,
 the integrals corresponding to the component diagrams in Figure~\ref{fig:fig2} diverge.
As we said,
the standard procedure at this point is to introduce a regulator in order to make finite each of
the component diagrams. We shall, instead, proceed differently and show that it is indeed possible to
have all integrals finite at each step of the calculation, so that no regularization is actually
necessary.
We shall first illustrate the method on the example of $L=3$. In the next Section  we shall
extend it iteratively to general $L$.

\begin{figure} [ht]
%[!htb]
% [!htbp]
 %\begin{minipage}[t]{\linewidth}

  \vskip -90pt
    \centering
    \includegraphics[width=1\linewidth]{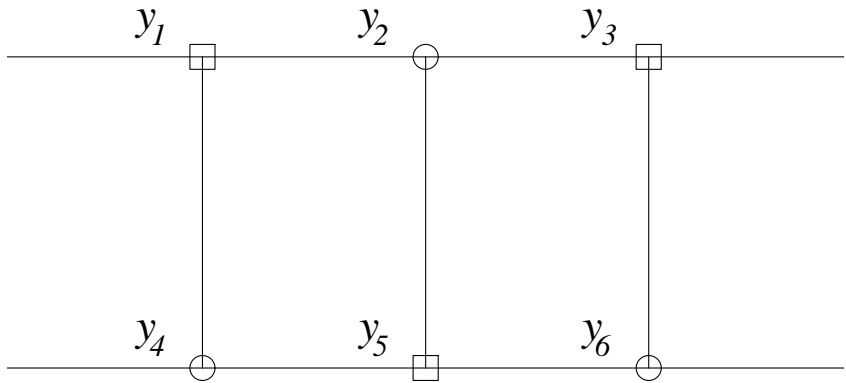}
  \vskip -50pt
  %   \vskip -8cm
\caption{The two possible sets of integration points for $L=3$.}
  \label{fig:fig3}
 %\end{minipage} \\ [2.5cm]
\end{figure}

Each of the diagrams in Figure~\ref{fig:fig2}
represents a 6-fold integral  over the internal integration points $y_1, \dots , y_6$ (see
Figure~3). Instead of trying to compute the 6-fold integral for each diagram and then summing
the results, we split the integration into two 3-fold integrals:
one set over integration variables with odd indices $\{y_{\rm odd}\}$, and a second
set over integration variables with even indices $\{y_{\rm even}\}$
according to the chess-board like scheme of Figure~\ref{fig:fig3}.
%%%%%%%%%%%%%%%%%%%%%%%
As a result we can represent the 4-point function $G_3$ in the form
\be
G_3(x_1,x_2,x_3,x_4) = \int d^4 \{y_{\rm even}\} \; W_3(x_1,x_2,x_3,x_4,\{{y_{\rm even}}\}) \, ,
\label{G_3int1}
\ee
where $W_3$ is given by an expression similar to  eq.~(\ref{G_3}) where only integrations over $\{y_{\rm odd}\}$ appear.
We now show that all the integrations over $\{y_{\rm odd}\}$  can be explicitly
carried out  in each of
the diagrams $D_1, \dots , D_6$ of Figure~\ref{fig:fig2} and that they are all finite. Putting all the resulting terms  together
one obtains an expression for $W_3$, which can be integrated also over $\{y_{\rm even}\}$ giving a finite result.

Note that one can exchange the order of the integrations over $\{y_{\rm even}\}$ and over $\{y_{\rm odd}\}$.
This leads to two apparently very different representations for the same 4-point correlation function $G_L$
(see bellow).
Their equivalence is a manifestation of the conformal invariance of $G_L$.

As can be noted by inspection, in each of the integration points  in the diagrams of Figure~\ref{fig:fig2} there is either a
$\delta$-function coming from the  propagator of the auxiliary field $F$, or two fermion propagators and a scalar
propagator.
The integration in the first case is trivial.
To perform the integration in the second case  we use the identity
\be
\widehat \de_1 \; \widehat \de_2 \int d^4 x_0 { 1 \over x^2_{10} x^2_{20} x^2_{30}} =
-4 \int d^4 x_0 { \widehat{ x_{10} } \; \widehat{ x_{02} } \over x^4_{10} x^4_{20} x^2_{30}} =
 - 4 \pi^2 { \widehat{ x_{13} } \; \widehat{ x_{32} } \over x^2_{13} x^2_{23} x^2_{12}} \, ,
\label{papa}
\ee
where $\widehat{ x_{ij} } = \sigma_{\mu} x_{ij}^{\mu}$. This identity follows by applying
the operator $\widehat \de_1  \widehat \de_2$ to  the
explicit expression of the integral in the l.h.s.\  whose evaluation can be
found in~\cite{davyd1}. Eq.~(\ref{papa}) is valid only in 4 space-time dimensions.

\begin{figure} [ht]
%[!htb]
% [!htbp]
 %\begin{minipage}[t]{\linewidth}

  \vskip -120pt
    \centering
    \includegraphics[width=1\linewidth]{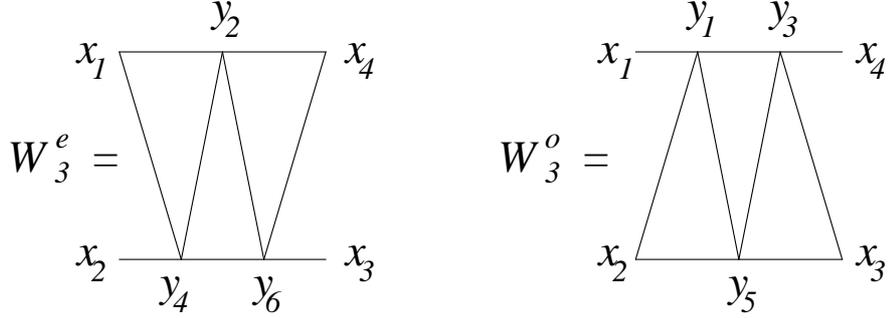}
  \vskip -40pt
  %   \vskip -8cm
\caption{The two functions entering eq.~(\ref{G3_final}) for the $L=3$ case.}
  \label{fig:fig4}
 %\end{minipage} \\ [2.5cm]
\end{figure}

Due to the chess-board choice of $\{y_{\rm even}\}$ and $\{y_{\rm odd}\}$, the application of eq.~(\ref{papa})
in any of the integration points does not alter the expressions involving the other integration points from the same set.
The same is also  true when a  $\delta$-function from the $F$ propagator is involved.
Hence,  one can always perform the integrations over {\it all} the points belonging to  one of the two sets.
Choosing, for example, to integrate first over $\{y_{\rm odd}\}$, we reduce
the expression of $G_3$ to the form in eq.~(\ref{G_3int1}), where the function $W_3$ is a linear combination
of products of propagators with numerators containing traces
of products of (up to 6)  $\widehat{ x_{ij} }$ factors. Each term in $W_3$ separately
contains high powers of $x^2$ in the denominator and diverges if integrated over $\{y_{\rm even}\}$.
However, after factoring out a common denominator and simplifying the numerator, it turns out that
in the whole function, due to cancellations, there are only simple $1/x^2$ poles
and the remaining $\{y_{\rm even}\}$ integrations display no divergencies. The final expression is
\ba
G_3 &=& - \; x_{14}^2 \; \int d^4y_2 d^4y_4 d^4y_6 \; W_3^e(x_1,x_2,x_3,x_4; y_2,y_4,y_6)  \nonumber \\
& = & - \; x_{23}^2 \; \int d^4y_1 d^4y_3 d^4y_5 \; W_3^o(x_1,x_2,x_3,x_4; y_1,y_3,y_5) \ ,
\label{G3_final}
\ea
where the second identity follows by interchanging the order of the
integrations over  $\{y_{\rm even}\}$ and $\{y_{\rm odd}\}$.
The functions $W_3^e$ and $W_3^o$
are given  by the effective diagrammatic representation of  Figure~\ref{fig:fig4}, where all the lines
represent massless scalar propagators.
%%%%%%%%%%%%%%%%%%%%%%%

Since in each internal vertex in Figure~\ref{fig:fig4} enter exactly 4 propagators, the integrals in eq.~(\ref{G3_final}) are
conformal, with conformal weights of the external
legs  equal to one.
This means that the correlator   can be expressed as a function of the two cross-ratio variables
\be
r= {x_{13}^2 x_{24}^2 \over x_{12}^2 x_{34}^2 } \ , \qquad s= {x_{14}^2 x_{23}^2 \over x_{12}^2 x_{34}^2 } \, ,
\label{cross-ratios}
\ee
in the form
\be
G_3(x_1,x_2,x_3,x_4) = {1 \over x_{12}^2 x_{34}^2 } F_3(r,s) \, .
\label{G3 conformal}
\ee
The integral in eq.~(\ref{G3_final}) has been considered, in a particular kinematical regime appropriate for
taking the on-shell limit, in~\cite{volovich} where also a Mellin--Barnes representation has been derived.

Before proceeding to the proof in the case of general $L$, we list the expressions
for the simpler  cases of $L=1$
\be
G_1(x_1,x_2,x_3,x_4) = - \int {d^4x_5 \over x_{15}^2 x_{25}^2 x_{35}^2 x_{45}^2 }
= -{\pi^2 \over x_{12}^2 x_{34}^2} \;  \Phi^{(1)}(r,s) \, ,
\label{1loop}
\ee
and $L=2$,
\ba
G_2(x_1,x_2,x_3,x_4) & = & x_{14}^2 \int {d^4x_5 d^4x_6 \over x_{15}^2 x_{16}^2 x_{26}^2 x_{35}^2 x_{45}^2 x_{46}^2 x_{56}^2}
\nonumber \\
 & = & x_{23}^2 \int {d^4x_5 d^4x_6 \over x_{15}^2 x_{25}^2 x_{26}^2 x_{35}^2 x_{36}^2 x_{46}^2 x_{56}^2} \nonumber \\
 && \nonumber \\
& = & {\pi^4 \over x_{14}^2 x_{23}^2} \;  \Phi^{(2)} \left( {r \over s} , {1 \over s} \right) \, \ .
\label{2loop}
\ea
Interestingly, they both reduce to the
massless $\varphi^3$ ladder, extensively studied in the literature~\cite{davyd1,davyd2,Broadhurst}.
The functions $\Phi^{(1)}$ and $\Phi^{(2)}$
 are the first two members of an infinite family of conformal integrals introduced  in~\cite{davyd2}.
The equivalence of the two representations for the  $L=2$ case has been emphasized in~\cite{magic} and used to prove
the equality of apparently different conformal integrals.

\section{All order result}
\label{sec:AOR}

In principle one can proceed as discussed in the previous section also in the case of higher $L$ ladder diagrams.
However, both the number of component diagrams as well as the complexity of each diagram grow very fast.
Here we shall use a different approach, namely we shall derive recursively  the general expression for
the massless ladder superdiagram with four external scalar legs $\varphi$ and $L$ rungs.
Expanding in component fields only the superpropagator in the leftmost rung one obtains a Bethe--Salpeter (BS) like
recursive equation relating the scalar ladder with $L+1$ rungs $G_{L+1}$ to  ladders with $L$ rungs, with
4 scalars ($G_L$), 2 fermions and 2 scalars ($G_{L}^{\psi}$), 2 auxiliary fields $F$ and 2 scalars
($G_{L}^{F}$), as external legs. Precisely we get the relation
\ba
G_{L+1}(x_1,x_2,x_3,x_4) &=& - \int d^4x_5 d^4x_6  \;
{ \tr \left[ {\widehat x_{65}} \; G_{L}^{\psi}(x_5,x_6,x_3,x_4) \right] \over x_{56}^4 x_{16}^2 x_{25}^2} \;
 \nonumber  \\
 &&  \nonumber  \\
&& - \int d^4x_5 d^4x_6 \; { \delta(x_{56}) \over x_{16}^2 x_{25}^2} \;
G_{L}(x_5,x_6,x_3,x_4) \nonumber \\
&&  \nonumber  \\
&& + \int d^4x_5 d^4x_6 \;  { 1 \over x_{16}^2 x_{25}^2 x_{56}^2} \;
G_{L}^{F}(x_5,x_6,x_3,x_4) \ ,
\label{BS scalar1}
\ea
where the ordering of the points corresponds to the convention of Figure~\ref{fig:fig1}.
The external fields in $x_3$ and $x_4$ are always scalars,  the
external fields in $x_5$ and $x_6$ are scalars in $G_L$,
fermions  in $G_{L}^{\psi}$ and  auxiliary fields in $G_{L}^{F}$.
The minus sign in the first term is due to the presence of an extra closed fermion loop,
while the minus sign in the second term
is due to the $(-1)$ sign in the $F$ propagator.
After integrating the $\delta$-function in the second term and simplifying  $G_{L}^{F}$, one obtains
\ba
\!\!\!\! G_{L+1}(x_1,x_2,x_3,x_4) &=& - \int d^4x_5 d^4x_6  \;
{ \tr \left[ {\widehat x_{65}} \; G_{L}^{\psi}(x_5,x_6,x_3,x_4) \right] \over  x_{56}^4 x_{16}^2 x_{25}^2}  \;
 \nonumber  \\
 &&  \nonumber  \\
\!\!\!\! && - \int d^4x_5  \; { 1 \over x_{15}^2 x_{25}^2} \;
G_{L}(x_5,x_5,x_3,x_4) \nonumber \\
 &&  \nonumber  \\
\!\!\!\! && + \int d^4x_5 d^4x_6   { 1 \over x_{16}^2 x_{25}^2 x_{56}^4} \;
G_{L-1}(x_6,x_5,x_3,x_4) \; .
\label{BS scalar2}
\ea
In this form the relation contains both  $L$ and $L-1$
four scalar ladders, as well as the $L$-rung ladder diagram with two fermion and two scalar external lines.
For the latter correlator, by expanding the leftmost rung, we derive the BS equation
\ba
G_{L+1}^{\psi}(x_1,x_2,x_3,x_4) &=&  \int d^4x_5 d^4x_6  \;
 { {\widehat x_{16}} \; {\widehat x_{65}} \; {\widehat x_{52}} \over x_{16}^4  x_{56}^4 x_{25}^4} \;
G_{L}(x_5,x_6,x_3,x_4) \nonumber  \\
 &&  \nonumber  \\
&& + \int d^4x_5 d^4x_6
 { {\widehat x_{16}}
 {\widetilde  {G_{L}^{\psi}}(x_5,x_6,x_3,x_4)}
   {\widehat x_{52}} \over x_{16}^4  x_{56}^2 x_{25}^4} \; ,
\label{BS fermion}
\ea
where ${\widetilde  {G_{L}^{\psi}}}$ is obtained from  ${G_{L}^{\psi}}$ by inverting the direction of the fermion
line, which amounts to the substitution $ {\widehat x_{ij}} \rightarrow {\widehat x_{ji}}$,
$ {\widehat x_{ij}} \; {\widehat x_{jk}} \; {\widehat x_{kl}}
\rightarrow {\widehat x_{lk}} \;  {\widehat x_{kj}} \; {\widehat x_{ji}}$ etc.

We shall first present the solutions of these equations, then we shall derive them.
As already mentioned, the ladder superdiagram with $L$ rungs is planar for odd $L$, and twisted
for even $L$. In order to work with only planar drawings we shall treat separately the two cases of
even and odd $L$.

\begin{figure} [ht]
%[!htb]
% [!htbp]
 %\begin{minipage}[t]{\linewidth}

  \vskip -70pt
    \centering
    \includegraphics[width=1\linewidth]{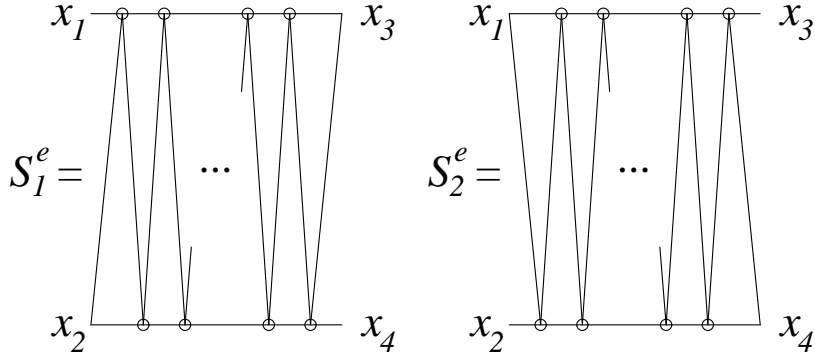}
  \vskip -70pt
  %   \vskip -8cm
\caption{The case of even $L$.}
  \label{fig:fig5}
 %\end{minipage} \\ [2.5cm]
\end{figure}

For the four scalars ladder and even values of $L$ we find
\ba
G_{L \; {\rm even}}(x_1,x_2,x_3,x_4) &=&  x_{23}^2  \; S_1^e(L ; x_1,x_2,x_3,x_4) \;
\nonumber \\
& = &  x_{14}^2 \;  S_2^e(L ; x_1,x_2,x_3,x_4)  \ .
\label{GL scalar even}
\ea
The functions
 $S_1^e$ and $S_2^e$ are diagrammatically depicted  in Figure~\ref{fig:fig5}, where all the lines
represent massless scalar propagators, and the circles denote the
$L$ integration points.
%%%%%%%%%%%%%%%%%%%%%%%

\begin{figure} [ht]
%[!htb]
% [!htbp]
 %\begin{minipage}[t]{\linewidth}

  \vskip -70pt
    \centering
    \includegraphics[width=1\linewidth]{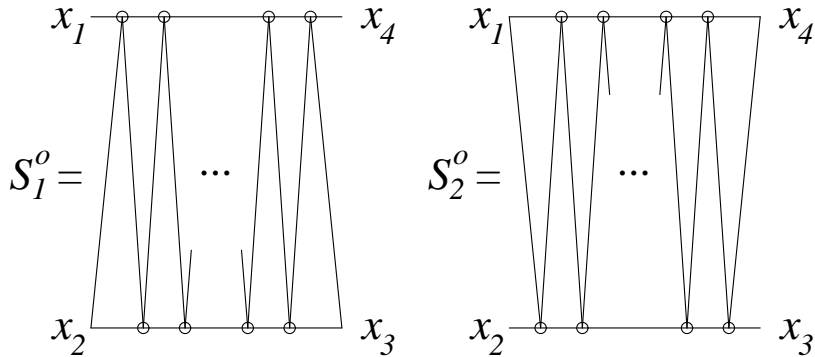}
  \vskip -70pt
  %   \vskip -8cm
\caption{The case of odd $L$.}
  \label{fig:fig6}
 %\end{minipage} \\ [2.5cm]
\end{figure}

For the four scalars ladder and odd values of $L$ we find
\ba
G_{L \; {\rm odd}}(x_1,x_2,x_3,x_4) &=& - \;  x_{23}^2  \; S_1^o(L ; x_1,x_2,x_3,x_4) \;
\nonumber \\
& = &  - \; x_{14}^2 \;  S_2^o(L ; x_1,x_2,x_3,x_4)  \ ,
\label{GL scalar odd}
\ea
where $S_1^o$ and $S_2^o$ are depicted in Figure~\ref{fig:fig6}.
%%%%%%%%%%%%%%%%%%%%%%%

In the case  of two fermions and
two scalars in the external lines one finds for $L$ even
\be
G_{L \; {\rm even}}^{\psi}(x_1,x_2,x_3,x_4) = - \int d^4x_5
{  {\widehat x_{14}} \; {\widehat x_{45}} \; {\widehat x_{52}} \over x_{15}^2 x_{25}^4 } \;
S_1^o(L-1 ; x_1,x_5,x_4,x_3) \; ,
\label{GL fermion even}
\ee
while for $L$ odd
\be
G_{L \; {\rm odd}}^{\psi}(x_1,x_2,x_3,x_4) =  \; \int d^4x_5
{  {\widehat x_{14}} \; {\widehat x_{45}} \; {\widehat x_{52}} \over x_{15}^2 x_{25}^4 } \;
S_1^e(L-1 ; x_1,x_5,x_4,x_3) \; .
\label{GL fermion odd}
\ee

By a direct  calculation one can check  that the results for the $L=1$, $L=2$ and $L=3$ ladders
are given by the expressions in eqs.~(\ref{GL scalar even}) to~(\ref{GL fermion odd}).
We now prove the general formulae by induction, using the BS equations~(\ref{BS scalar2})  and~(\ref{BS fermion}).
Assume that eqs.~(\ref{GL scalar even}) to~(\ref{GL fermion odd}) hold up to a given $L$ and substitute them in
the r.h.s of eq.~(\ref{BS scalar2})
(we take for definiteness $L$ to be odd).
With the help of eq.~(\ref{papa}) we can perform one of the integrations in the term coming from
$G_{L}^{\psi}$, simplify the trace of four $\sigma$ matrices in the numerator
and use the identity
\be
 (x_{ij}.x_{kl}) = {1 \over 2 } \;  (x_{il}^2+x_{jk}^2-x_{ik}^2-x_{jl}^2) \
\label{prod to square}
\ee
to express the resulting scalar products as squares of coordinate differences.
Putting together all the terms, the potentially dangerous
$1/x_{56}^4$ behaviour in the last line of eq.~(\ref{BS scalar2}) disappears, and we get the result in
 eq.~(\ref{GL scalar even}). The other equations~(\ref{GL scalar odd}) to~(\ref{GL fermion odd})
are proven in a similar way.

The equality of the two different representations for $G_L$ in eq.~(\ref{GL scalar even}) and eq.~(\ref{GL scalar odd})
 is again a
manifestation of the conformal invariance of these expressions. Actually, for any $L$,
one can write
\be
G_L(x_1,x_2,x_3,x_4) = {1 \over x_{12}^2 x_{34}^2 } F_L(r,s) \, ,
\label{GL conformal}
\ee
where $r$ and $s$ are the cross-ratios defined in eq.~(\ref{cross-ratios}).
This representation makes also manifest  the invariance of the ladder diagram under the simultaneous
exchanges  ($x_1 \leftrightarrow x_4$ and $x_2 \leftrightarrow x_3$);
 ($x_1 \leftrightarrow x_3$ and $x_2 \leftrightarrow x_4$);
($x_1 \leftrightarrow x_2$ and $x_3 \leftrightarrow x_4$).

 \section{Alternative representations of $G_L$}
\label{sec:ARG}

One can use the conformal invariance of the function $G_L$ eq.~(\ref{GL conformal}) to simplify
the integral which has to be computed. In particular, without loss of generality one can
perform a special conformal transformation to send
one of the points (say $x_4$) to infinity. Setting
\be
f_L (x_1,x_2,x_3) \equiv
{\lim_{ x_4\rightarrow \infty}} \; x_4^2 \; G_L(x_1,x_2,x_3,x_4) = {1 \over x_{12}^2 } \;
F_L \left({x_{13}^2  \over x_{12}^2 }, {x_{23}^2 \over  x_{12}^2 } \right) \, ,
\label{GL_limit}
\ee
the calculation is reduced to that of an effective 3-point function.
Indeed the $x_4$ dependence of $G_L$ can be reconstructed unambiguously from
the knowledge of $f_L (x_1,x_2,x_3)$.

\begin{figure} [ht]
%[!htb]
% [!htbp]
 %\begin{minipage}[t]{\linewidth}

  \vskip -110pt
    \centering
    \includegraphics[width=1\linewidth]{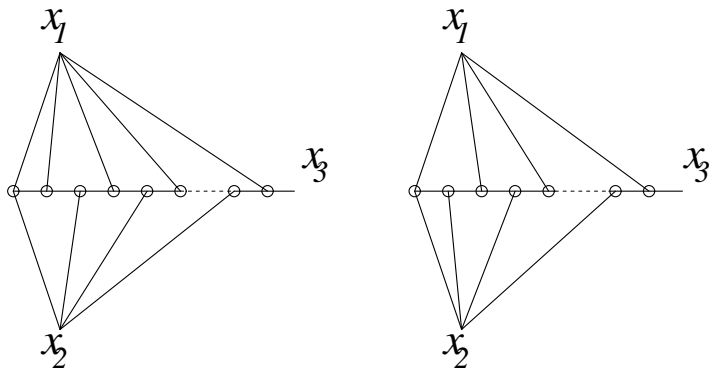}
  \vskip -40pt
  %   \vskip -8cm
\caption{Representation for the 3-point function $f_L (x_1,x_2,x_3)$.
The left picture corresponds to  $L$ even,
 and the right one to $L$ odd. For $L$ odd there is an overall minus sign
 which is not shown in the picture.}
  \label{fig:fig7}
 %\end{minipage} \\ [2.5cm]
\end{figure}

Taking the limit $x_4\rightarrow \infty$ has a simple graphical representation,
namely it corresponds to erasing from the diagram all lines connecting $x_4$ with the other points.
This explains how, combining the $x_4 \rightarrow \infty$ limit  with
repeated application of the equality of the two rows of eqs.~(\ref{GL scalar even}) and~(\ref{GL scalar odd})
(with the purpose of connecting as many as possible lines with the point $x_4$), we can reduce significantly the
number of propagators in the diagram.
Applying this procedure one can derive for $f_L (x_1,x_2,x_3)$ the representation
 depicted in Figure~\ref{fig:fig7}, where the left picture corresponds to  $L$ even,
 and the right one to $L$ odd.
 The diagrams in Figure~\ref{fig:fig7} are obtained by
 systematically replacing
 the representation involving $S_1^e$ with the one involving $S_2^e$
 and the representation involving $S_1^o$ with the one involving $S_2^o$,
 for all subfunctions $G_{K}(x_1,x_2,y,x_4)$ for $K=L-1,L-2,\dots,2$,
 thus at each step increasing by one the number of propagators ending in $x_4$.
 Finally one sends $x_4$ to infinity.
%%%%%%%%%%%%%%%%%%%%%%%

Other expressions of the same function can be obtained  either by using a different
sequence of identities, or by sending some other point to infinity.
The one we have singled out has two remarkable properties. On the one hand
it involves the minimal possible number of propagators \ie  $2L+1$,
 with an integrand having a numerator
equal to one.
On the other hand it has a form where exactly 3 propagators are attached to each
integration point.
Hence, the functions in Figure~\ref{fig:fig7} are  nothing but particular 3-point functions of composite operators in the
massless $\varphi^3$  theory. Our analysis shows that these 3-point functions are finite.

In Figure~\ref{fig:fig7} it is manifest
that the functions $f_L (x_1,x_2,x_3)$ defined in eq.~(\ref{GL_limit})
satisfy a recursive relation, namely
\be
f_L (x_1,x_2,x_3) = - \int {d^4x_0 \over x_{10}^2 x_{30}^2 } f_{L-1} (x_2,x_1,x_0)
 \ ,
\label{int_eq1}
\ee
and the differential equation
\be
\Box_{x_3} f_L (x_1,x_2,x_3) = {4 \pi^2 \over x_{13}^2 } f_{L-1} (x_2,x_1,x_3)
 \ ,
\label{diff_eq1}
\ee
which in turn implies a differential recursive relation for the functions $F_L(r,s)$.

\begin{figure} [ht]
%[!htb]
% [!htbp]
 %\begin{minipage}[t]{\linewidth}

  \vskip -100pt
    \centering
    \includegraphics[width=1\linewidth]{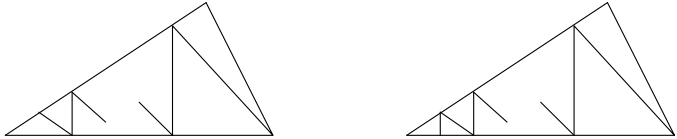}
  \vskip -110pt
  %   \vskip -8cm
\caption{Dual representation for the 3-point function $f_L (x_1,x_2,x_3)$.
The left picture corresponds to  $L$ even,
 and the right one to $L$ odd.}
  \label{fig:fig8}
 %\end{minipage} \\ [2.5cm]
\end{figure}

One can  also rewrite the integrals for $f_L$ as dual
$L$-loop ``momentum space like'' integrals as shown in Figure~\ref{fig:fig8}.
If one takes all the ``momenta'' to be incoming,
then the one in the left vertex is $x_{12}$, the one in the right vertex is $x_{23}$,
the one in the upper vertex is $x_{31}$, while those in the intermediate lines
are constrained by ``momentum'' conservation holding  in each vertex.
Again the left picture corresponds to  $L$ even, and the right one to $L$ odd.
Similar representations can be derived also for
the 4-point functions, (see \eg~\cite{volovich} for the case $L=3$).
Let us stress that, although all the propagators are massless,
since in general $x_{ij}^2 \neq 0$,  one has to compute  these ``momentum space like'' integrals
with off-shell external legs.
%%%%%%%%%%%%%%%%%%%%%%

Finally, we have to put back the overall power of $4 \pi^2$ which we suppressed
for simplicity during the calculation. As already mentioned, the ladder diagram with $L$ rungs
contains $3L+2$ propagators, each proportional to $1/4 \pi^2$.
On the other hand, in deriving eqs.~(\ref{GL scalar even}) and~(\ref{GL scalar odd}) we have applied  $L$ times
the relation in  eq.~(\ref{papa}), producing a factor $(4 \pi^2)^{L}$.
Hence, the overall multiplicative factor in $G_L$ and $f_L$ will be
$1/(4 \pi^2)^{2L+2}$.

\section{The 3-loop integral}
\label{sec:3LP}

Note that the cross ratios $r$ and $s$ defined in eq.~(\ref{cross-ratios}) cannot take arbitrary values.
In particular, in the Euclidean regime they satisfy the constraints
\be
r \geq 0 \ , \quad s  \geq 0 \ , \quad
1+r^2+s^2-4 r - 4 s - 4 r s \leq 0 \ .
\label{rs constraints}
\ee
The last inequality is saturated when  the 4 points $x_1, \dots ,x_4$ can
be mapped by a conformal transformation to lie on a straight line,
which, after sending by a special conformal transformation $x_4$ to infinity,
reduces to the condition that $x_1$, $x_2$  and $x_3$ lie on a straight line.
In this Section we shall compute the 3-loop integral
\be
I_3(x_1,x_2,x_3)  = \int { d^4x_5 d^4x_6 d^4x_7 \over
x_{15}^2 x_{25}^2 x_{26}^2 x_{17}^2 x_{37}^2 x_{56}^2 x_{67}^2}
\label{f3_1d}
\ee
in this special kinematical regime~\footnote{The result for the general case will be reported in a separate
paper~\cite{inprep}.}.
The precise relation between the integral $I_3$ and the function $f_3$ is
\be
I_3(x_1,x_2,x_3) = - (4 \pi^2)^8 f_3(x_1,x_2,x_3) \ .
\label{f3_1da}
\ee

Note that, since  $I_3(x_1,x_2,x_3)$ has singularities only at coinciding
arguments, $x_i=x_j$, the knowledge of $I_3$ in this special configuration is sufficient to completely
determine its singular behaviour.  We use translation invariance to set $x_2=0$
and dilatation invariance to set $x_3 = {\bf 1}$,
with {\bf 1} a fixed unit vector.  Since we have chosen the three points  to lie on a line,
the  point $x_1$ can be parameterized as $x_1 = u {\bf 1}$.
One has to treat separately the three cases, $u<0$, $0<u<1$ and $u>1$.
We first consider $0<u<1$.
In this representation the limit $u\rightarrow 0$ corresponds to
$x_{12}\rightarrow 0$, while  the limit $u\rightarrow 1$ corresponds to
$x_{13}\rightarrow 0$. We shall analyze the limit $x_{23}\rightarrow 0$,
which is related to $u \rightarrow \infty$, later.

In order to compute $I_3(u  {\bf 1}, 0 ,  {\bf 1})$, we  use eq.~(\ref{int_eq1})
to relate $f_3$ to $f_2$, eqs.~(\ref{GL_limit}) and~(\ref{2loop}) to
express $f_2$ in terms of $\Phi^{(2)}$ and
the parametric integral representation~\cite{davyd1}
\be
 \Phi^{(2)} \left( x,y \right) =  -{1 \over 2} \int_0^1 d t \;
 {{\rm ln}(t) \left(  {\rm ln}(t) + {\rm ln}({y \over x}) \right)
 \left( 2 \; {\rm ln}(t) + {\rm ln}({y \over x})  \right)
 \over  y t^2 + (1-x-y) t +x }
 \; .
\label{Phi2_2}
\ee
Putting everything together, one gets
\ba
&& \!\!\!\!\!\!\!\!\!\!\!\!\!\!\!\!\!\!
I_3(u  {\bf 1}, 0 ,  {\bf 1}) = % - {1 \over u^2} \; F_3\left({(u-1)^2 \over u^2}, {1 \over u^2}\right) =
 \nonumber \\
&&  \nonumber \\
&&  \!\!\!\!\!\!\!\!\!\!\!\!\!\!\!\!\!\! = \; - {\pi^4  \over 2} \int_0^1 d t \; \int d^4 x_0
 {{\rm ln}(t) \left( {\rm ln}(t)+ {\rm ln}(u^2/x_0^2) \right)
 \left(  2 \; {\rm ln}(t) + {\rm ln}(u^2/x_0^2)  \right)
 \over  (x_0-{\bf 1})^2  (x_0-u {\bf 1})^2 (x_0-t u {\bf 1})^2} \; .
\label{1Dim_int}
\ea
In the integral over $x_0$ one can pass to spherical coordinates, such that $(x_0. {\bf 1}) = R \; {\rm cos}(\theta)$
and perform first the integral over  $\theta$ and then the radial integral in $R$.
Taking into account also the
trivial integration over the remaining two  angles, the result is
\ba
 && I_3(u  {\bf 1}, 0 ,  {\bf 1})  = { \pi^6 \over u(1-u)} \times
 \nonumber \\
&&  \nonumber \\
&&
\left(
2 \int_0^1 { dt \ln(t) \polylog_3(t u) \over (1-t)}
-3 \int_0^1 {dt \ln(t)^2 \polylog_2(t u) \over (1-t)}
\right. \nonumber \\
&&  \nonumber \\
&&
-2 \int _0^1 {dt \ln(t)^3 \ln(1-t u) \over (1-t)}
-2 \ln(u) \int_0^1 {dt \ln(t) \polylog_2(t u) \over (1-t)}
 \nonumber \\
 &&  \nonumber \\
&&
-3 \ln(u) \int_0^1 { dt \ln(t)^2 \ln(1-t u) \over (1-t)}
 -2 \ln(u)^2 \int_0^1 { dt \ln(t) \ln(1-t u) \over (1-t)} \nonumber \\
% &&  \nonumber \\
&& +12 \; \polylog_4(u) \left( \ln(1-u) - \ln(u) \right)                       \nonumber \\
% &&  \nonumber \\
&&    -2 \; \polylog_3(u) \left( \polylog_2(u)  +3 \ln(u)( \ln(1-u)  -  \ln(u))
 \right)  \nonumber \\
%  &&  \nonumber \\
&& +  \;  \polylog_2(u) \ln(u) \left(  2 \; \polylog_2(u)
-4/3 \; \ln(u)^2   +2 \ln(u) \ln(1-u) \right)
    \nonumber \\
%     &&  \nonumber \\
&& + \pi^2 / 3 \; \left( \polylog_3(u)  -  \polylog_2(u) \ln(u) - \ln(u)^2 \ln(1-u) \right)  \nonumber \\
% &&  \nonumber \\
&& \left.  +2 \;  \zeta(3) \left( \polylog_2(u) + 3 \ln(u) \ln(1-u) \right)  -2 \; \pi^4 /15  \; \ln(1-u) \right)
 \; ,
\label{1Dim_final}
\ea
where ${\rm Li}_n(z)=\sum_{k=1}^{\infty} z^k/ k^n$  are the index $n$ polylogarithms.
One can rewrite eq.~(\ref{1Dim_final}) in terms of
harmonic polylogarithms~\cite{remiddi} of weights up to 5.
However, the supersymmetric massless ladder  for $L \geq 3$,
unlike the massless $\varphi^3$ ladder~\cite{davyd2},
cannot be expressed in a closed form in terms of usual polylogarithms $\polylog_n$.

From eq.~(\ref{1Dim_final}) one can compute the leading singular behaviour of the function
in the limit  $u \rightarrow 0$ (corresponding to $x_{12}\rightarrow 0$)
\be
\lim_{u \rightarrow 0} I_3(u  {\bf 1}, 0 ,  {\bf 1}) =
\left( -{4\over 3} \ln(u)^3 + 8 \ln(u)^2 - 20 \ln(u)+20 -4 \; \zeta(3) \right)  \pi^6 \ ,
\label{limit0}
\ee
and in the limit $u \rightarrow 1$ (corresponding to $x_{13}\rightarrow0$)
\be
\lim_{u \rightarrow 1} I_3(u  {\bf 1}, 0 ,  {\bf 1}) = 12 \;  \zeta(3) \; (1-\ln(1-u))  \;  \pi^6
\ .
\label {limit1}
\ee

To compute the limit $x_{23}\rightarrow 0$, we find it more convenient to
use a different parametrization, namely to choose $x_1$ = {\bf 1},
with {\bf 1} a fixed unit vector
 and $x_3 = \tilde u  {\bf 1}$. For non-vanishing $u$ and $\tilde u$
 the two parametrizations can be related  by a scale transformation
 and the  substitution $\tilde u= 1/u$.
Using again eqs.~(\ref{int_eq1}), (\ref{GL_limit}), (\ref{2loop})
and~(\ref{Phi2_2})
we get
\ba
&& \!\!\!\!\!\!\!\!\!\!\!\!\!\!\!\!\!\!
I_3( {\bf 1}, 0 , \tilde u  {\bf 1}) =  % - F_3((\tilde u-1)^2, \tilde u^2)   =
\nonumber \\
&&  \nonumber \\
&& \!\!\!\!\!\!\!\!\!\!\!\!\!\!\!\!\!\!
= - {\pi^4 \over 2} \int_0^1 d t \; \int d^4 x_0
 {{\rm ln}(t) \left( {\rm ln}(t) -   {\rm ln}(x_0^2) \right)
 \left(  2 \; {\rm ln}(t) -{\rm ln}(x_0^2)  \right)
 \over  (x_0-{\bf 1})^2  (x_0-t {\bf 1})^2 (x_0- \tilde u {\bf 1})^2}
 \; .
\label{1Dim_int2}
\ea
In this expression one can take the limit $\tilde u \rightarrow 0$ under the integral,
obtaining the finite value
\be
\lim_{\tilde u \rightarrow 0} I_3(  {\bf 1}, 0 , \tilde u {\bf 1}) = 20 \; \zeta(5) \;  \pi^6
\ .
\label{limitinf}
\ee
This is not a surprise, since in this limit the 4-point function $G_3$
reduces to a finite 3-point function. Actually this is true to all orders,
hence all functions $G_L$ will be finite in the limit $x_{23}\rightarrow 0$.
The finite 3-loop 3-point function gives rise to
a logarithmic divergence in the 4-loop propagator.
Our result, eq.~(\ref{limitinf}), is in perfect agreement with
the calculation of this propagator correction~\cite{kazakov}.

To summarize, the 3-loop integral $I_3(x_1,x_2,x_3)$
has a cubic logarithmic divergence for $x_{12} \rightarrow 0 $,
a linear logarithmic divergence for $x_{13} \rightarrow 0 $
and has a finite limit for $x_{23} \rightarrow 0 $.

\section{Conclusions and outlook}
\label{sec:CONO}

In this paper we have worked out a method to substantially simplify the computation of the supersymmetric
massless ladder 4-point diagrams with four external scalar legs at
arbitrary loop order. The main result of our paper are eqs.~(\ref{GL scalar even}) and~(\ref{GL scalar odd}) which
 express the ladder 4-point diagram with $L$ rungs, $G_L$, in terms of
conformal integrals.
For $L \geq 3$ the integrals in the supersymmetric case are different from the ones
in the non supersymmetric case.
We obtained two different representations for the conformal integrals
in the supersymmetric case as 3-point functions and as dual momentum space like integrals.
We  derived the expression
for the 3-loop integral $I_3(x_1,x_2,x_3)$
in the special case when the three points $x_1$, $x_2$  and $x_3$ lie on a straight line
and we computed the singularities of the function.
The resulting expression (see eq.~(\ref{1Dim_final})) turns out to be rather involved.
For a generic configuration of $x_1$, $x_2$  and $x_3$   it is,
not unexpectedly, even more complicated~\cite{inprep}.

In principle, following the same steps as in Section~\ref{sec:3LP} one can iterate the
procedure and derive $f_4$ from $f_3$ etc.. However the complexity of the
corresponding formulae grows very fast. Thus effectively such a direct approach seems very
difficult, if not impossible, to pursue. We believe that a  better understanding of the
 structure and the properties
of the relevant conformal integrals is necessary at this point.
Another interesting open problem is how to extend our construction to diagrams involving also
vector superfields.

\section*{Acknowledgements}

\noindent
It is a pleasure to thank Emery Sokatchev for
discussions.
This work was supported in
part by the INTAS grant 03-516346,
MIUR-COFIN 2003-023852, NATO PST.CLG.978785, the RTN grants
MRTNCT-2004-503369, EU MRTN-CT-2004-512194, MRTN-CT-2004-005104.
Partial support from MIUR-Italy (PRIN06) is also gratefully acknowledged.

\end{document}